\documentclass[useAMS,usenatbib]{mnras}

\usepackage[T1]{fontenc}
\usepackage{ae,aecompl}

\usepackage{graphicx}
\usepackage{float}
\usepackage{url}
\usepackage{braket}
\usepackage{gensymb}
\usepackage{commath}
\usepackage{amsmath}	% Advanced maths commands
\usepackage{amssymb}	% Extra maths symbols
\usepackage{mathtools}	% Extra maths symbols
\usepackage{wrapfig}
\title[A census of kpc-scale cold-gas outflows in Seyfert 2s]{This is not the feedback you have been looking for: nearby optical AGN rarely drive kpc-scale cold-gas outflows}
  \author[Nedelchev et al.]{Borislav
  Nedelchev$^{1,2}$\thanks{E-mail: b.nedelchev@herts.ac.uk}, Marc
  Sarzi$^{1}$ and Sugata Kaviraj$^{1}$\\ $^{1}$Center for Astrophysics
  Research, University of Hertfordshire, College Lane, AL10 9AB
  Hatfield, UK \\
  $^{2}$European Southern Observatory, Karl-Schwarzchild-Str. 2, 85748, Garching bei M$\ddot{u}$nchen, Germany}

\begin{document}

% % % % % % % % % % % % % % % HEADER

\maketitle

\label{firstpage}

% % % % % % % % % % % % % % % % % % % % % % % % % % % % % % % % % % % % % % % % % % % %

%%%%% DEFINE FIGURES WITH NEW COMMANDS %%%%%
\newcommand{\kms}{$\rm km\,s^{-1}$}
\newcommand{\NaD}{Na$\,$D}

%%%%%%%%%%%%%%%%%%%%%%%%%%%%%%%%%%%%%%%%%%%%%%%%
%\newcommand{\kms}{\,km\,s$^{-1}$}

% % % % % % % % % % % % % %  ABSTRACT

\begin{abstract}

We study the interstellar Na\textsc{\,i} $\lambda \lambda 5890, 5895$ (\NaD) absorption-line doublet in a nearly-complete sample of $\sim$9900 nearby Seyfert 2 galaxies, in order to quantify the significance of optical AGN activity in driving kpc-scale outflows that can quench star formation. Comparison to a carefully matched sample of $\sim$44,000 control objects indicates that the Seyfert and control population have similar \NaD\ detection rates ($\sim 5-6 \%$). Only 53 Seyferts (or 0.5\% of the population) are found to potentially display galactic-scale winds, compared to 0.8\% of the control galaxies. While nearly a third of the \NaD\ outflows observed in our Seyfert 2 galaxies occur around the brightest AGN, both radio and infrared data indicate that star formation could play the dominant role in driving cold-gas outflows in an even higher fraction of the \NaD-outflowing Seyfert 2s. Our results indicate that galactic-scale outflows at low redshift are no more frequent in Seyferts than they are in their non-active counterparts, that optical AGN are not significant contributors to the quenching of star formation in the nearby Universe, and that star-formation may actually be the principal driver of outflows even in systems that do host an AGN. 

\end{abstract}


% % % % % % % % % % % % % % % % % % % % % % % % % % % % % % % % % % % % % % % % % % % % % % % % % % % 

\newcommand{\IncludeFigOne}{
\begin{figure*}
\vspace{1cm}
\centering \includegraphics[width=0.89\linewidth ]{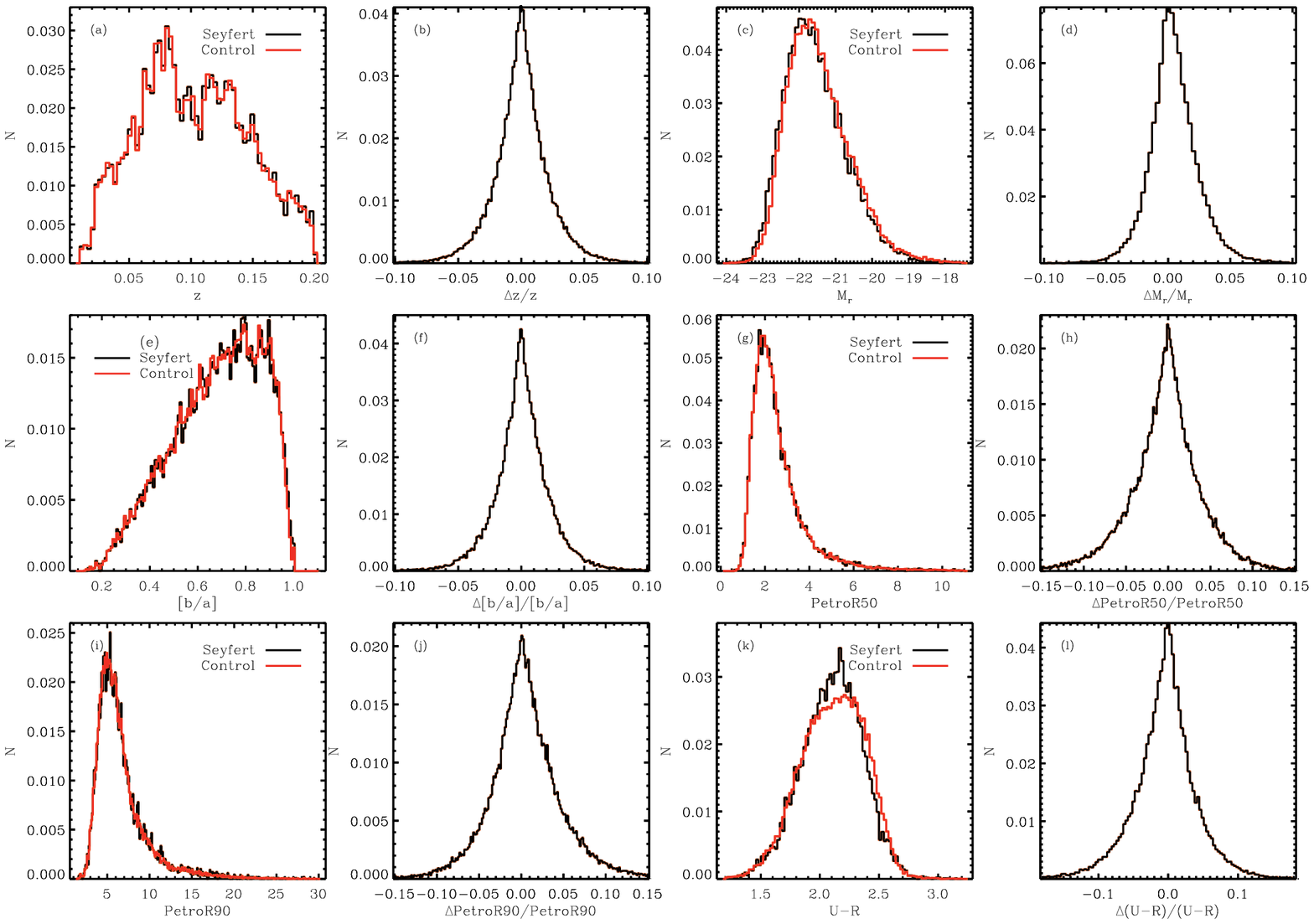}
\caption{The result of our control matching procedure. Panels (a),
  (c), (e), (g), (i), (k) show the distribution of our Seyfert 2 and
  control samples in all parameters used in the matching process -
  redshift (z), absolute r-band magnitude ($M_{r}$), isophotal semi-
  minor and major ratio $([b/a])$, Petrosian radius containing 50\% of
  the light ($PetroR50$), Petrosian radius containing 90\% of the
  light ($PetroR90$), and the difference in the absolute magnitude in
  the u- and r- bands ($U-R$). The remaining panels (b), (d), (f), (h),
  (j), (l) present the fractional difference between the
 aforementioned parameters for the galaxies in the control sample and
  the Seyfert 2 galaxy they match, normalised by the value of the
  corresponding parameter for the Seyfert 2. We achieve a good overall
  matching between the Seyfert 2 and control galaxy distribution, with the slight exception of $U-R$ where our matching procedure
  produces slight differences near the peak of the colour distribution.}
\label{fig:cnt} 
\end{figure*} 
} 

\newcommand{\IncludeFigTwo}{
\begin{figure*}
\centering 
\includegraphics[angle=-90,width=0.99\linewidth]{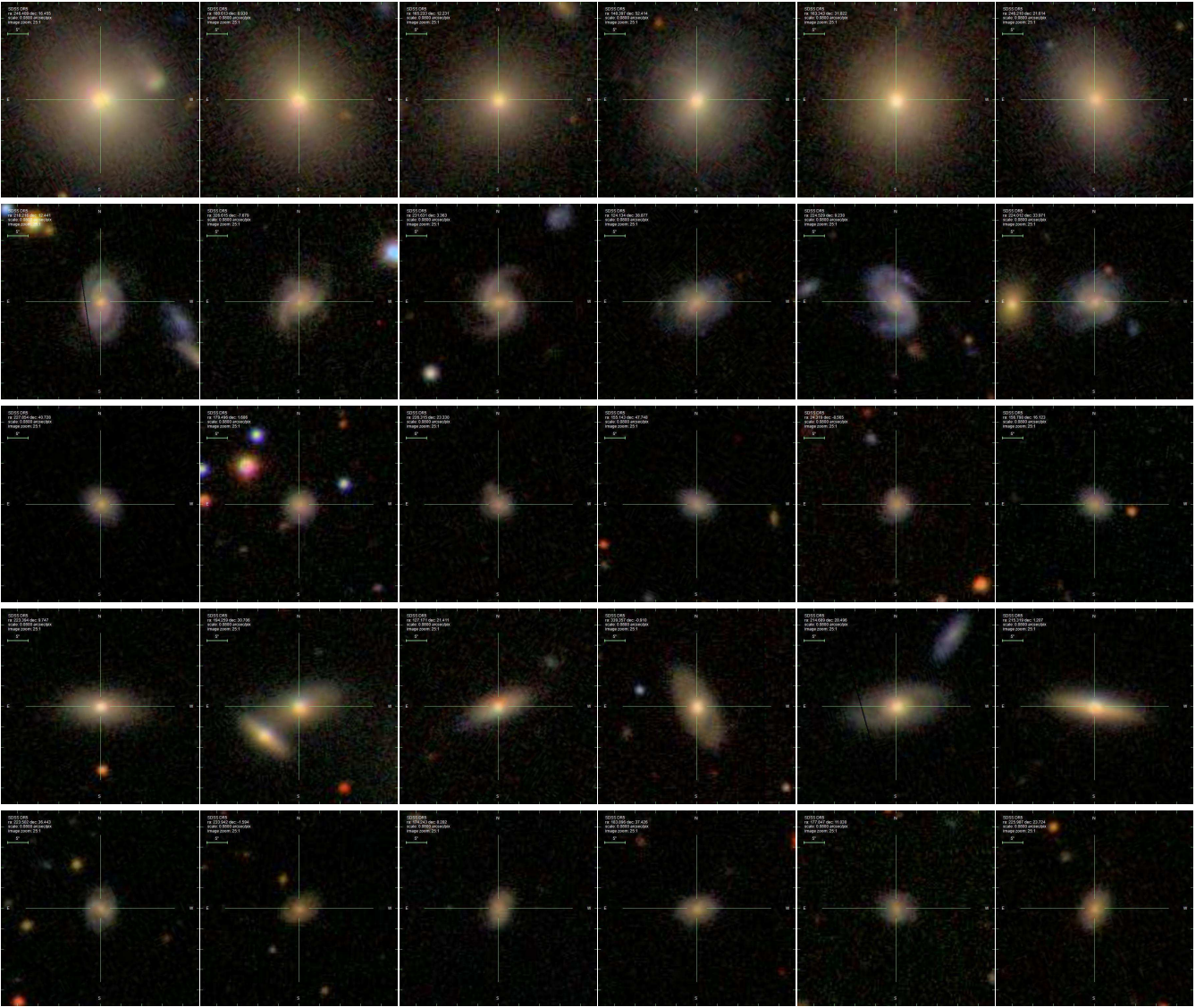} 
 \vspace{2cm}
\caption{SDSS colour images for Seyfert and control galaxies in our study. The image in the first column is the Seyfert 2 galaxy in question, while the next four columns in the corresponding row are the five closest-matched control galaxies. The top two rows show morphologically matched early- and late-type objects,
  respectively. All other rows show the results of our
  matching procedure when morphological criteria cannot be used due to an ``uncertain'' classification in Galaxy Zoo for the Seyfert in question.}

\label{fig:cntGZ} 
\end{figure*} 
} 

\newcommand{\IncludeFigThree}{
\begin{figure*}
\vspace{1cm}
\centering 
\includegraphics[width=0.89\linewidth,keepaspectratio]{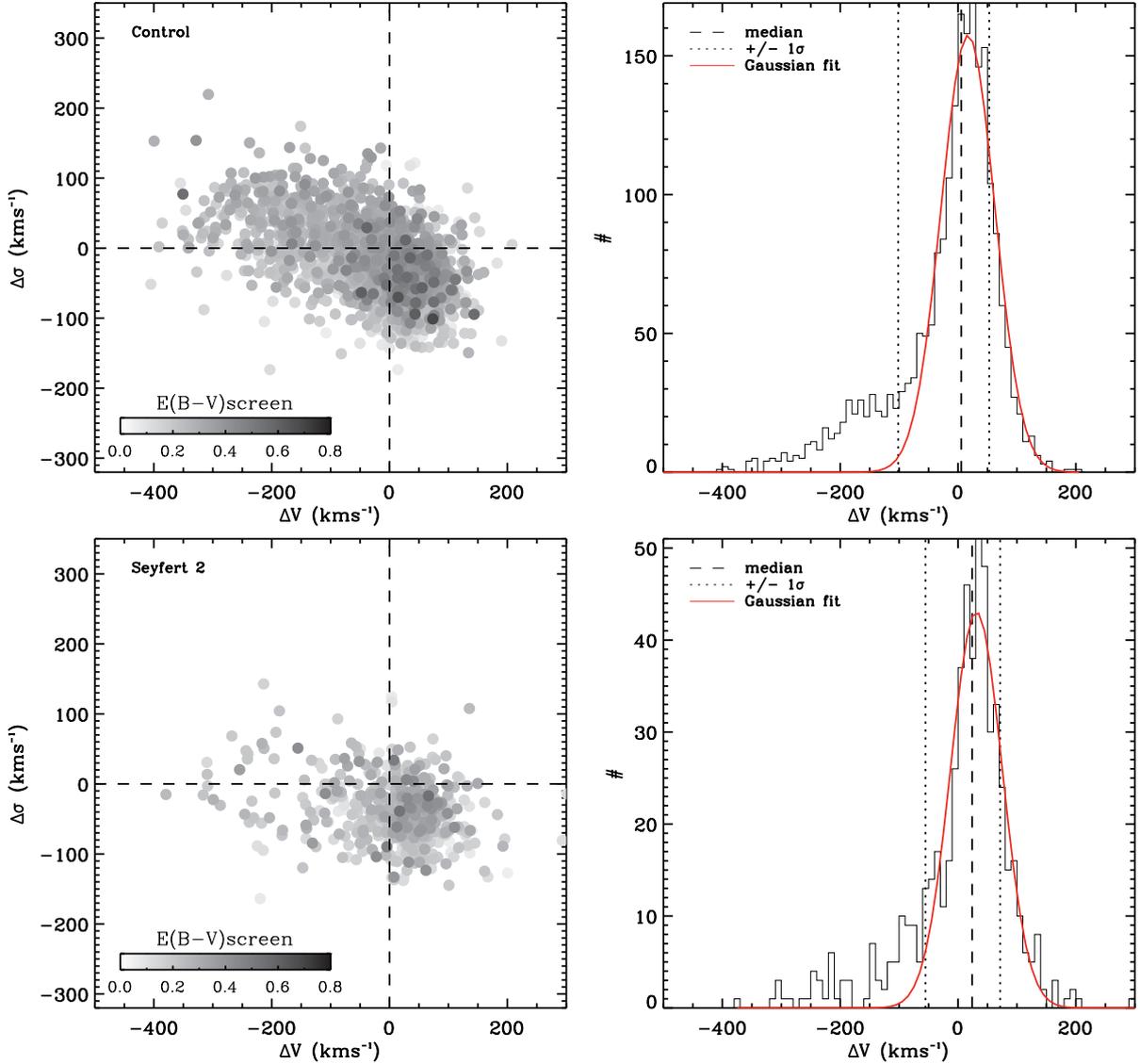}
\caption{Kinematics of the \NaD\ absorption
  component relative to the systemic properties of the galaxy, for our control and Seyfert 2 sample, where the observed \NaD\ excess can be safely attributed to interstellar absorption.
  \textbf{Top left:} $V_{\rm Na\, D}-V_{\star}$ ($\Delta V$) versus $\sigma_{\rm Na\,
    D}-\sigma_{\star}$ ($\Delta \sigma$) for our control sample,
  colour-coded by the dust reddening $E(B-V)$. Only galaxies with $A/N >4$ for the \NaD\ absorption and with $E(B-V) > 0.05$ are shown (see text for more details).
  \textbf{Top right:} Histogram showing the distribution of the velocity offsets $\Delta V =
  V_{\rm Na\, D}-V_{\star}$ for all the control galaxies shown in
  the top-left panel. The dashed line indicates the median of the
  distribution, and the dotted lines correspond to the $\pm 1 \sigma$
  levels containing $68\%$ of the $\Delta V$ distribution. The
  red line shows the best-fitting Gaussian to the underlying
  distribution and emphasises the pronounced tail of galaxies with a
  $\Delta V < -100$.
  \textbf{Bottom left:} An identical $\Delta V$ vs. $\Delta \sigma$ diagram as in the top-left panels, but now for our Seyfert 2 sample.
  \textbf{Bottom right:} An identical $\Delta V$ distributions as in the top-right panels, but now for our Seyfert 2 sample.}
\label{fig:main} 
\end{figure*} 
} 

\newcommand{\IncludeFigFour}{
\begin{figure}
\centering 
\includegraphics[width=0.79\linewidth,keepaspectratio]{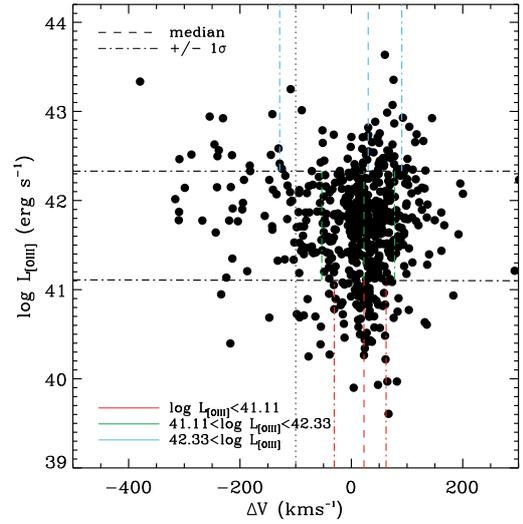}
\caption{Dependence of the velocity shift $\Delta V = V_{\rm Na\,
    D}-V_{\star}$ on the AGN bolometric luminosity, as traced by
  $\log{L_{\rm [\textsc{O\,iii}]}}$, which, in turn, is derived from 
  the extinction-corrected [\textsc{O\,iii}] flux provided by the
  OSSY catalogue.
%
% for all  Seyfert 2 where the excess \NaD\ profile was attibuted to ISM.
%
  The horizontal dashed-dotted lines indicate the ($\pm 1 \sigma$)
  16\% and 84\% percentiles of the $\log{L_{\rm [\textsc{O\,iii}]}}$
  distribution. In each of the three $\log{L_{\rm [\textsc{O\,iii}]}}$
  regions defined by these $\pm 1 \sigma$ $\log{L_{\rm
      [\textsc{O\,iii}]}}$ limits, the vertical dashed and
  dashed-dotted lines show the median and the $\pm1\sigma$ values for
  the velocity shifts $\Delta V$, respectively.
  Finally, the gray vertical dotted lines indicate our $\Delta V
  =-100$ \kms\ threshold for identifying \NaD\ outflows, as derived
  from our control sample (see text in \S~\ref{subsec:resC}).
  Only at the highest $\log{L_{\rm [\textsc{O\,iii}]}}$ end does the
  $\Delta V$ distribution become significantly skewed, with a tail of
  outflowing objects. These objects comprise the most luminous $\sim20\%$ of Seyfert 2 galaxies, and 32\% of all \NaD-outflowing Seyfert 2 objects.
%
% within the three corresponding $\log{L_{\rm [OIII]}}$ bins in red for
% the $38<\log{L_{\rm [OIII]}}<40.9$ bin, green for the
%$40.9<\log{L_{\rm [OIII]}}<42.1$ interval, and blue for the
%$42.1<\log{L_{\rm [OIII]}}<44$ range.
%
}
\label{fig:L_AGNN} 
\end{figure} 
}

\newcommand{\IncludeFigFive}{
\begin{figure}
 \vspace{0.5cm}
\centering 
\includegraphics[width=0.79\linewidth,keepaspectratio]{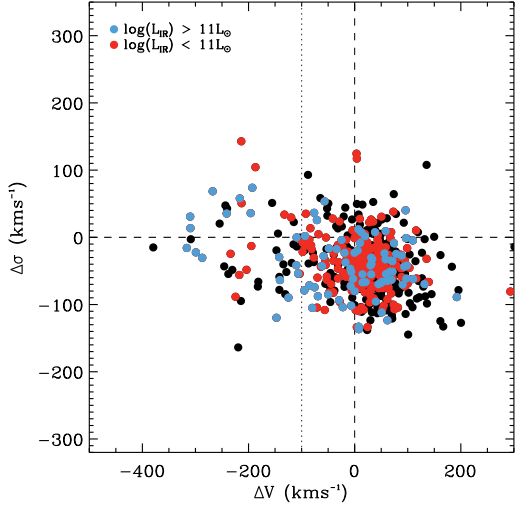}
\caption{$\Delta V$ vs. $\Delta \sigma$ for Seyfert 2 galaxies with interstellar \NaD\ absorption, similar to the one in
  Fig.~\ref{fig:main} but now showing the objects colour-coded with IR luminosity
  estimates from the catalogue of \citet{Elli16}. Blue and red
  points indicate galaxies with IR luminosities above
  and below the $L_{IR} = 10^{11}L_\odot$ threshold i.e. LIRGs and ULIRGs respectively. Among the outflowing Seyfert 2 galaxies (i.e. those to the left of the grey vertical dotted line that marks our $\Delta V =-100$ threshold for identifying
  \NaD\ outflows, see\S~\ref{subsec:resC}) 60\% would be classified as
  LIRGs or ULIRGs.}
\label{fig:IR_SY} 
\end{figure} 
} 

\newcommand{\IncludeFigSix}{
\begin{figure}
 \vspace{0.5cm}
\includegraphics[width=0.79\linewidth,keepaspectratio]{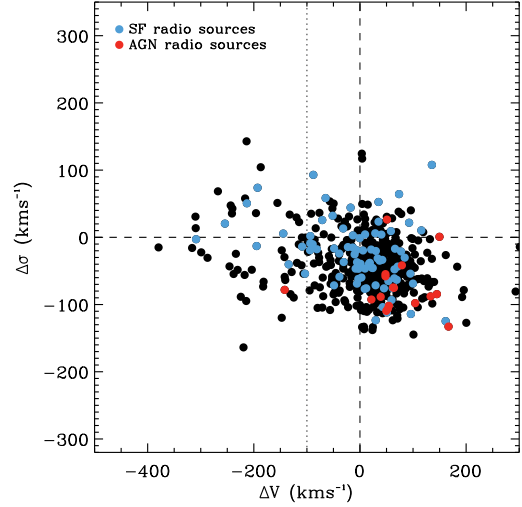}
\caption{$\Delta V$ vs. $\Delta \sigma$ for Seyfert 2 galaxies with 
  interstellar \NaD\ absorption, similar to the one
  Figs.~\ref{fig:main} and Fig.~\ref{fig:IR_SY}, but now showing radio sources that \citet{Bes12} associate with either AGN activity
  (red points) or star formation (blue points). Among the ten
  outflowing Seyfert 2 galaxies whose radio emission was classified in this way by
  \citeauthor{Bes12}, only one is considered by these authors as being
  powered by a central AGN.}
\label{fig:RAD_SY} 
\end{figure} 
} 

\newcommand{\IncludeTabone}{
\begin{table}
\centering
\renewcommand{\footnoterule}{}  % to avoid a line before footnotes
\begin{tabular}{llll}
\hline \hline

\textbf{Emission class} & \textbf{all} &  
\textbf{with \NaD} &  \textbf{\NaD\ outflows}\\
\hline
Seyfert 		& 9859 	& 561 (5.7\%) 	& 53 (9.4\%) \\
%Seyfert 		& 9859 	& 561 		 	& 53 \\
\hline
control 		& 44123 & 2174 (4.9\%) 	& 352 (16.2\%) \\
%control 		& 44123 & 2174 		 	& 352  \\
\hline
control-SF 		& 6710 	& 587 (8.7\%) 	& 92 (15.7\%)  \\
control-TO 		& 5895 	& 567 (9.6\%) 	& 147 (25.9\%)  \\
control-LINER 	& 859 	& 105 (12.2\%) 	& 27 (25.7\%)  \\
control-NE 		& 30659 & 915 (2.9\%) 	& 87 (9.5\%) \\
%control-SF 		& 6710 	& 587  	& 92  \\
%control-TO 		& 5895 	& 567  	& 147 \\
%control-LINER 	& 859 	& 105  	& 27  \\
%control-NE 		& 30659 & 915  	& 87  \\
\hline
\end{tabular}
\caption[]{Breakdown of our Seyfert 2 and control samples, according to
  whether or not interstellar \NaD\ absorption was detected
  (third column, for objects with $A/N_{\rm Na\, D} > 4$ and $E(B-V) > 0.05$, see
  \S~\ref{sec:analysis}) and furthermore if it likely traces
  outflows (fourth column, for object with $\Delta V = V_{\rm Na\, D} - V_{\star} < -100$ \kms,
  see \S~\ref{subsec:resC}). Control galaxies are further subdivided
  according to their emission-line classification either as
  star-forming regions (SF), composite AGN/star-forming activity (TO),
  low-ionisation - possibly truly nuclear - emission regions (LI), or
  displaying little to no emission (NE). Percentages in the third column refer to the fraction of \NaD-outflowing objects with respect to the number of objects with detected \NaD\ interstellar absorption (third column).}
\label{tab:cnrl}
\end{table}
}
% % % % % % % % % % % % % % % % % % % % % % % %contor % % % % % % % % % %

% % % % % % % % % % % % DEFINE CITATION ALIASES ****************
\defcitealias{Oh11}{OSSY}

% % % % % % % % % % % % % % % % % % % % % % % % % % % % % % % % % % %

% % % % % % % % % % % % SET COUNTER for chapters to refer to 

\setcounter{secnumdepth}{3}

% % % % % % % % % % % % %

% % % % % % % % % % % % % INTRODUCTION ***********************

\section{INTRODUCTION}
\label{sec:intro}

Feedback from active galactic nuclei (AGN) is often invoked as a key mechanism for suppressing star formation in massive galaxies, where stellar winds and supernovae explosions are not sufficiently energetic to efficiently expel gas from their deep gravitational potential
wells \citep[e.g.][]{Sil98}. In such massive systems, the energy released when their supermassive black holes (SMBHs) accrete matter can exceed the gravitational binding energy of their host galaxies by several orders of magnitude. Thus, from a theoretical point of view, energetic feedback from AGN is thought capable of disrupting the progress of star formation in such systems, through the heating and ejection of the galaxy's gas reservoir \citep[][]{Fab12}. Such (negative) AGN feedback is frequently incorporated in models of galaxy formation, to prevent the overprediction of over-massive systems in the models and bring their predicted properties, like colours, in line with observations \citep[e.g.][]{Cro06,Sca12,Kav17}. 

From an empirical perspective, the AGN hypothesis is qualitatively supported by the observed correlations between SMBH masses and various global physical properties of their host galaxies (e.g velocity dispersion), which suggest that the host galaxy and its SMBH may grow in lockstep with each other \citep[see][for a review]{Kin15}. Nevertheless, the detailed processes behind this possible coevolution \citep{Kor13} remain unclear, and gaining insight into the impact that SMBHs can have on the evolution of their hosts in large statistical samples of galaxies is essential \citep{Scha12}.

The clearest examples of host-AGN interaction are arguably found in nearby brightest cluster galaxies. The AGN in these systems have been shown to deposit vast amounts of energy into the surrounding intracluster medium via heating and (mega-parsec scale) jets both observationally and by means of modeling \citep[e.g.][for a review]{Bin04, Scann05, Engl16, Git12}, which maintain the hot gas reservoirs in these systems, prevent cooling flows and thus suppress star formation \citep[e.g.][]{Bin95, Li15}. In lower-density environments (where the majority of galaxies live), the empirical picture is much less clear. Direct observational evidence for AGN feedback on galactic scales in such environments remains sparse. More specifically, while outflows have been detected in a number of instances around AGN in various gas phases, most of these detections have been made in ultra-luminous infrared galaxies or some of the closest quasar-host galaxies \cite[e.g.][]{Nyl13,Rup05b,Harr14}, with only a few examples where the AGN have been shown to couple with kpc-scale outflows that are capable of impacting star-formation on galactic scales \citep[for a review]{Cic14, Harr17}. Thus, it is still unclear to what extent the general AGN population could drive kpc-scale galactic outflows capable of limiting or quenching star-formation in the local Universe.

Indeed, some recent work has cast doubt on the ability of AGN (outside clusters) to strongly regulate star formation in the \textit{nearby} Universe. These studies indicate that the time delay between the peak of the star formation and the onset of AGN activity is several dynamical timescales \citep{Kav11,Kav15,Shab17}. Consequently, AGN couple mainly to residual gas, at a point where star formation has already declined and the original gas reservoir is already significantly depleted \citep{Sar16}. This, in turn, makes it unlikely that the AGN play a significant role in regulating their associated star formation episodes. A fuller understanding of the role of AGN in regulating star formation demands a direct study of whether outflows of neutral material (which are ultimately responsible for quenching star formation) are more likely launched in AGN hosts. Most importantly, a quantitative statement about the putative role of AGN in influencing the evolution of their host galaxy requires a study that employs a complete sample of such AGN in the local Universe. Performing such an analysis is the purpose of this paper. 

An efficient way to identify galaxies that are going through an outflow phase is to look for the presence of interstellar Na\textsc{\,i} $\lambda
\lambda 5890, 5895$ (\NaD) absorption that is blue-shifted with
respect to the systemic velocity of the galaxy. Indeed, the low-ionisation
potential of Sodium makes the detection of blue-shifted \NaD\ lines an
unambiguous signature of neutral interstellar material that is been
entrained within an outflow, observed against the stellar background of
the host galaxy \citep{Rup05a}. In this paper we measure the incidence and kinematics of interstellar \NaD\ absorption in one of the largest samples of nearby
Seyfert 2 galaxies drawn from the Sloan Digital Sky
Survey \cite{Aba09}, and compare these results with similar measurements
obtained in a carefully selected control sample designed to match our
Seyfert 2 galaxies in redshift, luminosity, size, light concentration,
apparent flattening, colour and, when available, morphological classification into early or late type objects. We use this comparison to quantify whether the frequency of kpc-scale cold-gas outflows increases in Seyfert 2's, with a view to understanding the impact that the central engines of Seyfert 2 galaxies have on their hosts.  
 
This paper is organized as follows. Section 2 gives a brief
description of our Seyfert 2 sample, and the procedure used to
construct the control sample. Section 3 describes how we identify and extract the kinematics
of \NaD\ absorption features. In Section 4 we present the
results of this analysis and in Section 5 we summarize our conclusions.

% % % % % % % % % % SAMPLE *********************
\IncludeFigOne
\IncludeFigTwo
\section{DATA AND SAMPLE SELECTION}
\label{sec:data_sample}

Both the analysis and the sample selection that underpins this work are based on Data Release 7 of the Sloan Digital Sky Survey \citep[SDSS;][]{Str02,Aba09}. For our measurements of \NaD\ line strength and position, we re-analyse the SDSS spectra that were used for the construction of the \citet{Oh11} value-added SDSS catalogue (hereafter OSSY), which analysed
in detail 664,187 galaxies at $0.0 < z < 0.2$ and provides key quantities such as the value of the stellar velocity dispersion and the strength of nebular emission lines. The relative intensity of emission lines, as measured in OSSY, was also used to select our core Seyfert 2 sample and classify the nebular emission of our control objects by means of standard BPT diagnostic diagrams \citep{Bal81}.
 
For measuring total luminosities and colours we obtain absolute,
de-reddened, K-corrected magnitudes from the SDSS DR7 in the u- and
r-band, that are based on the best-fitting de Vaucouleurs or exponential model
magnitudes ($ModelMag$). The K-corrections are taken from the
$kcorr$ value for the u- and r-band listed in the SDSS $Photoz$ table
\citep{OMil11}, whereas for computing absolute magnitudes we use the
luminosity distance returned by the $fCosmoDl$ functions in the SDSS
$CfunBASE$ library \citep{Tag10} and the SDSS redshift listed in the
$SpecObjAll$ table. All necessary isophotal measurements for measuring
the radial extent, flattening and concentration of the SDSS galaxies
are taken from the SDSS $PhotoObjAll$ table, in particular, the
r-band values for the $isoA$ and $isoB$ measurements of the major- and
minor-axis radii and the $petroR90$ and $petroR50$ entries for the
Petrosian radii enclosing 90\% and 50\% of the Petrosian flux.

Finally, in this work, we consider only the OSSY catalogue objects
that were visually inspected by the Galaxy Zoo citizen-science project
\citep{Lin11}. 
%
% with available de-biased ``super-clean''
% morphological classification for galaxies into either ellipticals or
% spirals, which were based on a vote agreement threshold of at least
% 80\%. 
%
Since around 94\% of the OSSY catalogue objects were covered by Galaxy Zoo, the requirement of a visually-classified morphology does not significantly reduce the final sample of 625,607 objects from which we select the Seyfert 2 and control samples that underpin this study. 

\subsection{Seyfert 2 sample}
\label{subsec:SYsample}

%As a starting point for selecting a sample of Seyfert 2 host galaxies, we look for objects in the OSSY catalogue which had emission-line amplitude-to-noise ratios (A/N) greater than 3, for both the [{\sc O$\,$iii}]$\lambda 5007$ and [{\sc N$\,$ii}]$\lambda 6584$ forbidden emission lines and for the H$\alpha$ and H$\beta$ recombination lines. This allows us to identify objects with nebular emission that is sufficiently strong to be classified using the BPT diagnostic diagram introduced by \citet{Vei87} that juxtapose the [{\sc N$\,$ii}]/H$\alpha$ and [{\sc O$\,$iii}]/H$\beta$ line ratios. 
%
%To separate galaxies with nuclear Seyfert activity (Sy) from those whose central nebular emission is dominated either by star-forming regions (SF), low-ionisation emission regions (LI, sometimes associated with true AGN activity) or a superposition of AGN and star-formation activity .(TO for ``transition''), we adopt the division lines from \citet{Kew01}, \citet{Kau03}, thus arriving at an initial sample of 10983 Seyfert galaxies. 

As a starting point for selecting a sample of Seyfert 2 host galaxies, we select objects in the OSSY catalogue which had emission-line amplitude-to-noise ratios (A/N) greater than 3, for both the [{\sc O$\,$iii}]$\lambda 5007$ and [{\sc N$\,$ii}]$\lambda 6584$ forbidden emission lines and for the H$\alpha$ and H$\beta$ recombination lines. Such a detection threshold allows us to safely place them on the so-called BPT \citep{Bal81} emission-line diagnostic diagrams and, more specifically, on the one introduced by \citet{Vei87} that combines the [{\sc N$\,$ii}]/H$\alpha$ and [{\sc O$\,$iii}]/H$\beta$ line ratios. We adopt the empirical demarcation of \citet{Kau03} to separate these galaxies into ones where the gas ionization is driven by star-formation (SF) activity and ones potentially driven by AGN. Furthermore, by applying the \citet{Kew01} maximum theoretical star-burst division line, we differentiate a class of galaxies with a superposition of AGN and star-formation activity (TO for ``transition''). Finally, the rest of the emission-line galaxies with plausible AGN contributions were divided on the basis of the empirical criterion developed in \citet{Sch07}, to isolate Seyfert galaxies (Sy) from the ones with low-ionization emission regions (LI), which can rarely be attributed to real AGN activity and are instead more often due to extended emission \citep{Sar10, Cid10, Cid11, Yan12, Singh13, Bel16}. 
%
%While the latter are sometimes attributed to real AGN activity, the LINER-like %emission in many of these systems is rather extended, challenging the view that it %is uniquely powered by an AGN \citep{Singh13}. 
%
As a result of this selection process we arrive at an initial sample of 10,983 Seyfert galaxies.
Since many of these Seyfert objects could potentially display broad-line regions (BLR), and a featureless AGN continuum that would considerably complicate our \NaD\ measurements \citep{Car17}, we further restrict ourselves solely to \textit{narrow-line} Seyfert 2 galaxies, by excluding the Type 1 Seyferts identified by \cite{Oh15}. This leaves us with a final sample of 9859 Seyfert galaxies.

\subsection{Control Sample}
\label{subsec:Csample}

In order to ascertain the importance of AGN in driving outflows, we will compare our \NaD\ analysis of the Seyfert 2 sample, to an identical analysis on a carefully selected sample of control galaxies. Following a procedure similar to
\citet{Wes07}, we proceed by finding, for each Seyfert 2 object in our sample, five control galaxies that are closest to the Seyfert in question in the following quantities:

\begin{enumerate}
\item Redshift, $z$, from the SDSS DR7 pipeline. 
\item Absolute r-band magnitude, $M_{r}$, K-corrected and de-reddened
  for the Galactic extinction, as described above. This is done to
  match galaxies as closely as possible in their total stellar
  luminosity.
\item Apparent flattening, $b/a$, using the r-band isophotal minor- to
  major-axis ratio. For spiral galaxies, this quantity provides a useful measure of inclination.
\item The radius containing 90\% of the Petrosian flux, $PetroR90$,
  which, when combined with our $z$ matching, enables us to match galaxies in their intrinsic size. This also ensures that the SDSS
  spectroscopic measurements encompass roughly the same fraction of
  galaxy light in our Seyfert 2 and control galaxies.
\item The radius containing 50\% of the Petrosian flux, $PetroR50$,
  which, when combined with our $PetroR90$ matching, allows us to match
  galaxies in their degree of light concentration. In the absence of a
  robust visual classification, this, to some extent, provides a proxy for the galaxy morphology \citep{Shi01,Str01}. 
\item $U-R$ color, based on their $M_{u}$ and $M_{r}$ absolute
  magnitudes. Combined with our $M_{r}$ matching, this allows us to pick
  objects with comparable positions in the $M_{r}$ vs. $U-R$
  color-magnitude diagram, therefore selecting galaxies with
  similar star-formation rates and star formation histories. 
\end{enumerate}

To find the best control objects for each of our Seyfert 2 galaxies we
decided against minimising the sum ($\Delta C$) of the absolute
differences in each matching parameter (e.g., $\Delta z$) as done
for instance in \citet{Wes07} and opted instead to consider such differences in relative terms (i.e. $\Delta z/z$) and  weighting them as follows:

\begin{multline}
\Delta C = \frac{\abs{\Delta z/z}}{0.05} + \frac{\abs{\Delta M_r/M_r}}{0.1} + 
\frac{\abs{\Delta [b/a]/[b/a]}}{0.05}  \\
+\frac{\abs{\Delta petroR90/petroR90}}{0.1}  
+\frac{\abs{\Delta petroR50/petroR50}}{0.05} \\
+\frac{\abs{\Delta (M_u-M_r)/(M_u-M_r)}}{0.1}
\end{multline}

This allows us to find matching control objects more precisely at the higher- and lower- end of the Seyfert 2 distribution for each of the previous parameters, and to assign a relative error meaning to our adopted weights. The weights themselves are chosen following an iterative process. Very broad initial constraints (i.e. asking all parameters to be not so precisely matched) were initially imposed and further tightened, guided by the specific number of available control galaxies for every Seyfert, in our parameters of interest, until our matching procedure achieved the relative best match across all six adopted parameters.

For each of our Seyfert 2 objects, we compute $\Delta C$ for all OSSY galaxies, and then pick the five objects with the lowest $\Delta C$ as control counterparts for the Seyfert in question. 
During this process, if a Seyfert 2 object had a robust
``super-clean'' Galaxy Zoo classification and was classified either as
a spiral or an elliptical, we proceeded to find the best five control
galaxies as done above, but while considering only OSSY objects that
were also robustly classified to be of the same morphological type. This selection process results in 44,123 unique control galaxies.
Fig.~\ref{fig:cnt} shows how our matching process returns control
galaxies, with parameter distributions that closely match the corresponding distributions for our Seyfert 2 sample (panels a, c, e, g, i, and k). Fig.~\ref{fig:cnt} also shows
how the values for these parameters for the five best control galaxies
are generally within 5 to 10\% of the ones of their parent Seyfert 2			
objects (panels b, d, f, h, j, and l), broadly consistent with the
weighting applied in Eq. (1). 
The only exception in this respect concerns our colour matching, since
the $U-R$ distributions for our Seyfert 2 and control sample differ slightly near their peaks and red tails.

Fig.~\ref{fig:cntGZ} further illustrates the quality of our control
sample selection, by presenting SDSS colour composite images for 5
randomly selected Seyfert 2 galaxies (left column) and their
respective five best controls (columns to the right). 
In Fig.~\ref{fig:cntGZ} the top two rows are Seyfert 2 galaxies that are classified by Galaxy Zoo as an elliptical and spiral respectively. 
Conversely, the last three rows show how our matching process returns control galaxies that look similar to our Seyfert 2 objects even when the morphology of the latter was classified as 'uncertain' by Galaxy Zoo. 

% % % % % % % % % % SAMPLE *********************
\section{Analysis}
\label{sec:analysis}

To assess the presence of interstellar \NaD\ absorption, and measure the
kinematics of these lines, we closely follow the procedure
described in \citet{Sar16}:

\begin{enumerate}
\renewcommand{\theenumi}{(\arabic{enumi})}

\item We normalise the SDSS spectra in the \NaD\ region using the
  best-fitting stellar continuum, obtained using the \texttt{pPXF} and
  \texttt{GandALF} spectral fitting codes \citep{Cap04, Sar06}, with the
  synthetic stellar population (SSP) of \citet{Bru03} used as stellar
  templates. This fitting is entirely consistent with the one carried
  out during the compilation of the OSSY catalogue, except that in this work
  we avoid using the semi-empirical templates that were constructed by
  \citet{Oh11} in order to better match the spectra of massive early-type 
  galaxies. 
%  
% This is done in order to compensate for template mismatch which is caused by the
% inherent limitations of the \citeauthor{Bru03} models in reproducing the spectra 
% of massive early-type galaxies (e.g., due to an overabundance of 
% $\alpha$-elements).
  Matching the nebular emission in this process is critical to account also for the
  possible presence of He$\,${\sc i} $\lambda 5875$ emission.

\item After normalising the SDSS spectra, we fit the resulting \NaD\
  absorption profile using the parametrisation given by \citet{Sat09}:

\begin{multline}
I(\lambda) = 1 - C_f
\Big\{ 1 - \exp \Big[
-2\tau_0 e^{-(\lambda -\lambda_{\rm blue})^2
%-\tau_{0,\rm blue} e^{-(\lambda -\lambda_{\rm blue})^2
/(\lambda_{\rm blue} b / c)^2} \\
- \tau_0 e^{-(\lambda - \lambda_{\rm red})^2
%- \tau_{0,\rm red} e^{-(\lambda - \lambda_{\rm red})^2
/(\lambda_{\rm red} b / c)^2}
\Big] \Big\}
\end{multline}

where $C_{f}$ is the covering factor of the absorbing cloud complex,
$\tau_{0}$ is the optical depth at the centre of the red \NaD\ line,
$\lambda_{blue}$, $\lambda_{red}$ are the red- or blue- shifted
wavelengths of the two \NaD\ lines, and $b_{D}=\sqrt{2}\sigma_{\rm Na\, D}$ is
the Doppler parameter. The red- or blue- shifted central-line values
and $b_{D}$ yield the quantity of interest here, namely the velocity
$V_{\rm Na\, D}$ and the width $\sigma_{\rm Na\, D}$ of the \NaD\ lines.
\end{enumerate}

The procedure mentioned above assumes a single velocity distribution
for the absorbing gas clouds along the line of sight\footnote{More
  specifically, it assumes a Maxwellian velocity distribution for the
  gas clouds and that the covering fraction $C_{f}$ is itself
  independent of the clouds' velocities.}. However, this may not be accurate for galaxies that host both a significant
``systemic'' population of absorbers settled in the galaxy plane and
a population of outflowing clouds, in particular, when looking at these objects from
an intermediate inclination angle \citep{Che10}.
While this approach would lead us to underestimate the outflow
velocity in galaxies that are viewed from intermediate inclination
angles, using a single \NaD\ velocity profile will suffice to capture 
the kinematic behaviour of the cold gas across tens of thousands of 
Seyfert 2 and control galaxies and, most noticeably, to estimate and compare 
the fractions of objects that could display cold-gas outflows in our samples.

As noted in \citet{Sar16}, one has to be aware that
after normalising the SDSS spectra, an \NaD\ absorption excess could
also be the result of template mismatch, for instance in the case of
objects with an enhanced [Na/Fe] abundance in their stellar population that is not accounted by our adopted SSP template
library \citep{Jeo13,Par15}. Fortunately, objects where the \NaD\ excess is entirely artificial (in the context of our procedure) can be isolated using the OSSY values for the $E(B-V)$ that affects the entire spectrum (as opposed to reddening that impacts just the nebular emission,
\citet{Jeo13}). In other words, very small values of $E(B-V)$ allow us to identify objects with little or no interstellar medium. 

To restrict our study to objects displaying an \NaD\ excess
that is truly due to interstellar medium (ISM) absorption, as opposed to
template-mismatch, we therefore select objects with $E(B-V) >
0.05$. This threshold appears to separate well the main
Gaussian distribution of $E(B-V)$ values across both our
Seyfert and control samples from the offset peak of quiescent
galaxies with zero or very small $E(B-V)$ values.

Finally, to define a detection threshold for an \NaD\ absorption excess
in the SDSS data, we run our \NaD\ fitting procedure through a large
number of simulated spectra, obtained by adding random noise and an
artificial \NaD\ absorption profile to the best-fitting stellar
population model. 
By exploring how well the input \NaD\ parameters and, in particular, $V_{\rm Na\, D}$ and $\sigma_{\rm Na\, D}$, are recovered as a
function of the observed $A/N_{\rm Na\, D}$ ratio between the peak amplitude of the \NaD\ profile and the input noise
level, we conclude that at least $A/N_{\rm Na\, D}>4$ values are
required to secure unbiased $V_{\rm Na\, D}$ and $\sigma_{\rm Na\, D}$
measurements. 
This is consistent with the $A/N>4$ threshold estimated by the
simulations of \citet{Sar06} for ionised-gas emission lines with Gaussian profiles (in their case for the 
[{\sc O$\,$iii}]$\lambda\lambda4959,5007$ doublet), which can be understood considering that the \NaD\ interstellar absorption profile can also be matched by means of two negative Gaussians of identical redshift and width in the low optical-depth regime where the \NaD\ absorption is weak and \NaD\ lines are far from saturating.

% % % % % % % % % % % % % % % % % % % % % RESULTS *************************

\IncludeTabone
\section{Results}
\label{sec:res}

We begin by discussing the incidence and kinematics of
\NaD\ interstellar absorption in our control sample, while also defining
the \NaD\ blue-shift threshold that identifies outflowing systems
(\S~\ref{subsec:resC}). We then compare these results with
what is observed in our Seyfert 2 sample (\S~\ref{subsec:resSY}).
However, before conclusions can be drawn from such a comparison, it is
important to first consider the incidence of interstellar \NaD\ absorption 
in both the control and Seyfert 2 samples. Tab.~\ref{tab:cnrl} shows that 
the fraction of Seyfert 2 objects where \NaD\ interstellar absorption
is detected (5.7\%) is consistent with that 
of control galaxies (4.9\%). 

Such a remarkable similarity, combined with our careful control
sample construction, allows us to directly compare the \NaD\
kinematics in our two samples and to reflect on the importance of AGN
feedback. 
In particular, this finding goes against the argument that the AGN in the Seyfert 2 galaxies would substantially ionise its surroundings, as reported for instance by \cite{Vil14} for their sample of bright type 2 quasars. The ionising action of the AGN would indeed reduce the incidence of NaD interstellar absorption among our Seyfert 2s and thus limit the usefulness of the \NaD\ lines as a tracer of cold-gas large-scale outflows only to objects with a faint AGN, which may not be able to provide much energetic feedback.

\subsection{Control Sample}
\label{subsec:resC}
\IncludeFigThree

Tab.~\ref{tab:cnrl} shows that the majority (69\%, or 30,659 out of 
44,128 objects) of our control galaxies exhibit no or very weak nebular 
emission, with most remaining objects being almost evenly split between 
star-forming (13\%) and composite (13\%) nebular emission, plus a minority
(3\%) of galaxies displaying LINER-like emission.
Tab. ~\ref{tab:cnrl} also shows that the incidence of interstellar Na
D absorption is around 9 -- 12\% for control galaxies with
detected ionized-gas emission, dropping to just 3\% among quiescent
objects, which is consistent with the weak or absent nebular emission in these systems. 

To understand the kinematics of interstellar \NaD\ absorption in our
control sample, we start by considering the top left panel of
Fig.~\ref{fig:main}, where the velocity $V_{\rm Na\, D}$ and velocity
dispersion $\sigma_{\rm Na\, D}$ of the \NaD\ absorption is compared to the galaxy's systemic velocity and central velocity dispersion ($V_{\star}$ and $\sigma_{\star}$ respectively. The colour-coding indicates the reddening by dust, $E(B-V)$.
In this $V_{\rm Na\, D}-V_{\star}$ versus $\sigma_{\rm Na\,
  D}-\sigma_{\star}$ diagram, we observe the same trend as that observed in \citet{Sar16}, in that a majority of objects with nearly zero (or even slightly positive) $V_{\rm Na\, D}-V_{\star}$ ($\Delta V$) and negative $\sigma_{\rm Na\,D}-\sigma_{\star}$ ($\Delta \sigma$) values is followed by a long tail of galaxies with increasingly blue-shifted and broad \NaD\ profiles, i.e. with progressively negative $\Delta V$ and
increasing $\Delta \sigma$ values. 

The bulk of the objects with $\Delta V\sim0$ consists of control
galaxies where the \NaD\ profile traces material that is likely settled
in a dusty disk which is dynamically colder than the stellar bulge (thus corresponding to $\Delta \sigma < 0$ values), or of
highly-inclined systems where it would be hard to detect possible
outflows. 
Conversely, detecting outflows and increasingly blue-shifted \NaD\
lines is facilitated by looking at more face-on galaxies, where broad \NaD\ profiles are also expected, as the line of sight intersects a multitude of cold-gas clouds moving at different velocities within large-scale bi-conical galactic winds \citep{Fuj09, Krum17}.
Visual inspection of our control galaxies combined with using the inclination values derived from the axis ratio of late-type galaxies (51\% of our control sample) confirms that the observed $\Delta \sigma$ vs $\Delta V$ anticorrelation is mainly driven by galaxy inclination. This trend is supported by the systematically higher $E(B-V)$ values among low-$\Delta V$ objects, as would be expected for nearly edge-on dusty systems \citep{Unt08,Mas10}.

We proceed to address when \NaD\ outflows are likely to occur in our
control galaxies, by studying the distribution of values for their \NaD\ velocity offset ($\Delta V$), as shown in the top right panel of Fig.~\ref{fig:main}.
The $\Delta V$ distribution is skewed toward
negative values (with a statistical skewness value of $\sim-1.42$) and has a pronounced tail of objects with blue-shifted \NaD\ interstellar
absorption profiles. Yet, as stated before, the bulk of the control
objects in Fig.~\ref{fig:main} shows $\Delta V\sim0$ values, so that a
Gaussian fit to the entire $\Delta V$ distribution captures the
distribution of the majority of these galaxies well, and allows us to isolate the tail of control objects that have \NaD\ outflows.  
Observing that the control galaxies with $\Delta V < -100$ \kms 
(the $\sim$16$^{\rm th}$ percentile of the overall $\Delta V$ distribution) 
already lie outside the region occupied by 90\% of the objects enclosed by the
fitted Gaussian (which has a FWHM of 45 \kms), we consider this $\Delta V$ threshold
to be a conservative estimate for the start of the tail of blue-shifted \NaD\
control objects. Consequently, we consider galaxies with $\Delta V < -100$ \kms\ 
as being likely to host cold-gas outflows, which are expected to be on kpc-scales  given that at the mean redshift of our sample ($z \sim 0.1$) the 3\arcsec\ SDSS fibers subtend a region 5.5kpc across. 

Among the control galaxies with detected \NaD\ interstellar absorption, 
around $\sim 16.2\%$ (Tab.~\ref{tab:cnrl}) display \NaD\ profiles
blue-shifted by more than 100 \kms, which amounts to $\sim 0.8\%$ of
the control sample and is consistent with the overall outflow detection rate found in \citet{Sat09}.
Interestingly, we find that \NaD\ outflows are not found most commonly in the star-forming objects (which host 16\% of outflowing objects). Instead, outflows seem to be produced more efficiently in galaxies with either central composite AGN/star-forming or LINER-like emission (in 25\% of the cases).
Outflows are also traced in quiescent control objects with
\NaD\ interstellar absorption (10\% of the cases) that generally
display weak emission and where reddening prevented, in particular, the detection 
of [{\sc O$\,$iii}]. 
As such, the upper-limits on the [{\sc O$\,$iii}]/H$\beta$ line ratio
indicates the presence of star formation or composite AGN/star-formation activity in most of these quiescent objects. 

To study the role of star formation in driving outflows and further ascertain when it most likely powers them, 
it is instructive to look into the infrared (IR) luminosity of our control sample. 
For instance, it is well established that more than 50\% of all luminous infrared galaxies
(LIRGs) or ultraluminous infrared galaxies (ULIRGs) have blue-shifted
\NaD\ profiles indicative of outflowing gas
\citep{Hec00,Rup02,Rup05a,Rup05b,Mar05}. 

To estimate the IR luminosities ($L_{IR}$) of our control galaxies with \NaD\
interstellar absorption, we employ the \citet{Elli16} catalog of
predicted infrared luminosities for galaxies in the SDSS DR7,
selecting only objects where the expected error in their infrared
luminosity estimate is less than 0.1 dex.
Among our control objects with interstellar Na D detection, 
1519 (70\%) have $L_{IR}$ values from \citeauthor{Elli16}, and around 
30\% (461) of these have $L_{IR} >10^{11} L_{\odot}$ that would 
identify them as LIRGs or even ULIRGs. 
The fraction of objects with $L_{IR} > 10^{11} L_{\odot}$ is
substantially higher for objects with \NaD\ outflows, being
around 55\% (148 objects out of 268), whereas for the bulk of our
control objects that have $\Delta V\sim0$ \kms\, the fraction of objects with $L_{IR} > 10^{11} L_{\odot}$ drops to 25\% (313 out of 1251).

A higher incidence of U/LIRGs amongst our \NaD-outflowing control
objects is consistent with many of them being powered by
star-formation, even though only 36 out of these 148 \NaD-outflowing
and IR-bright control objects are found in galaxies with star-forming
central regions. This may suggest either that circum-nuclear
star formation is driving the outflow or that, in many of these objects, shocks in the outflow lead to a more complex emission-line spectrum, as
found in \citet{Ho14}, which could be recognised as
LINER-like or composite AGN/star-forming emission (as found in 56\% of
the \NaD-outflowing and IR-bright objects). 
In fact, shocks may also help explain why, more generally, \NaD\ outflows are often associated with composite or LINER-like emission in our control galaxies, which can also be observed in the \NaD\ kinematic analysis of the \citet{Jeo13} sample by \citet[see their Fig. 4]{Par15}. 
  
\subsection{Seyfert 2 sample}
\label{subsec:resSY}

As noted above, the fractions of Seyfert
and Control objects with interstellar \NaD\ absorption are very
similar, which is important, as this allows us to compare the cold-gas
kinematics and, in particular, the incidence of \NaD\ outflows in our
two samples.

We proceed by studying the kinematics of the interstellar \NaD\ lines in
our Seyfert sample. The bottom left panel of Fig.~\ref{fig:main}, which shows
the same $\Delta V$ vs. $\Delta \sigma$ diagram that we previously discussed for our control galaxies, indicates that among our Seyfert sample there is also a population of galaxies with blue-shifted \NaD\ interstellar lines (e.g. with $\Delta V <0$) that are likely tracing outflowing material. 
The $\Delta V$ distribution for our Seyfert 2 sample (bottom right
panel of Fig.~\ref{fig:main}) shows a skewed distribution (with a statistical skewness $\sim -1.22$), similar to that observed in the control sample, with a tail of objects with $\Delta V < -100$ \kms\ extending away from the bulk of the systems that have nearly zero or slightly positive $\Delta V$ values.
Compared to the control sample, however, the tail of objects with
blue-shifted \NaD\ profiles is less pronounced, as it contains 9.4\% of
all Seyfert 2 galaxies with interstellar \NaD\ absorption (compared to
16.2\% for the control galaxies), corresponding to 0.53\% of the
overall Seyfert 2 sample (i.e. those with and without \NaD\ absorption). 
  
\IncludeFigFour

Given that the overall incidence of \NaD-outflowing systems is similar to that
found in our control sample, it is important to recall that our sample of
9859 Seyfert 2 galaxies offers a near complete view of the Seyfert 2
population within the footprint of the SDSS survey.
Finding only 53 of such objects with kpc-scale cold-gas outflows
already suggests that optical AGN activity cannot be important in
driving galactic winds and quenching star-formation in the nearby
Universe, even accounting for the fact that some more \NaD-outflowing
systems may have remained undetected in inclined Seyfert 2 galaxies.
In fact, the role of AGN feedback may be even smaller considering how
well connected Seyfert 2 and star-formation activity are
\citep[e.g.,][]{Kau03}, so that circumnuclear star formation may
also play a part in driving the \NaD\ outflows that we observe in our
Seyfert 2 sample.

To further quantify the importance of AGN feedback we
check if the radiative power output of the AGN, i.e. its bolometric
luminosity as traced by the extinction-corrected luminosity of the
[{\sc O$\,$iii}] emission line \citep{Lam09}, correlates with the
velocity offset $\Delta V$ of the interstellar \NaD\ absorption.
This is shown in Fig.~\ref{fig:L_AGNN}, where we find a modest
tendency for galaxies with higher values of [{\sc O$\,$iii}]
luminosity ($\log{L_{\rm [\textsc{O\,iii}]}}$) to display 
blue-shifted interstellar \NaD\ profiles (indicative of an outflow) 
more often. 
The fraction of \NaD-detected Seyfert 2 galaxies where $\Delta
V<-100$ \kms\ indeed increases from $\sim 3\%$ or $\sim 9\%$, 
for objects with low to intermediate [{\sc O$\,$iii}] luminosities, 
to $\sim 19\%$ for the most luminous Seyfert 2 nuclei.
Outflows are not predominantly found in such bright AGN (where we 
find $\sim32$\% of the outflows), however, and inclination biases 
can also be excluded as we do not find - at least for Seyfert 2 in 
late-type galaxies (which are 38\% of the Seyfert 2 hosts) - a 
correlation between axis ratio and [{\sc O$\,$iii}] luminosity.

\IncludeFigFive 
\IncludeFigSix 

If the weak trend shown by Fig.~\ref{fig:L_AGNN} could still be
interpreted as evidence for AGN feedback in our \NaD-detected Seyfert
2, one also has to keep in mind that the luminosity of the [{\sc
    O$\,$iii}] line in AGN is found to correlate with the
star-formation rate of the host galaxy \citep{LMas13}.

To check whether star-formation is possibly also driving the \NaD\
outflows observed in our Seyfert 2 sample, we draw again from the work
of \citet{Elli16} and look for the incidence of LIRGs or
ULIRGs among our Seyfert 2 galaxies.  
This is illustrated by Fig.~\ref{fig:IR_SY}, where we present the same
$\Delta V$ vs. $\Delta \sigma$ diagram for the \NaD-detected Seyfert 2s 
as in Fig.~\ref{fig:main}, but now highlighting objects with $L_{IR}$ 
above and below $L_{IR} = 10^{11} L_{\odot}$, which is the LIRG threshold. 
The \citeauthor{Elli16} catalogue provides $L_{IR}$ estimates for 308
(55\%) of our Seyferts with \NaD\ interstellar absorption, and of
these, around 28\% (88) are sufficiently IR-bright to be classified as
LIRGs or ULIRGs.
Among objects with $L_{IR}$ estimates, the fraction of \NaD-outflowing 
Seyferts (i.e. with $\Delta V< -100$ \kms) with $L_{IR} > 10^{11} L_{\odot}$ is substantially higher 
than for Seyferts that show little or no evidence of cold-gas outflows 
(i.e. with $\Delta V >-100$ \kms), with these fractions 
being 60\% (17 out of 28) and 25\% (71 out of 280), respectively.
The incidence of LIRGs or ULIRGs among \NaD-outflowing Seyfert 2s is 
thus nearly twice as that of bright AGNs (32\%), suggesting that 
circumnuclear star formation could indeed be driving many of the 
outflows observed among our Seyfert 2 galaxies. In fact, this possibility 
appears even more likely when considering that the fraction of IR-bright 
\NaD-outflowing Seyfert 2s is also remarkably consistent with the 
corresponding value for the \NaD-outflowing control objects (55\%, \S~\ref{subsec:resC}).

To further probe the relative role of AGN and star-formation in
driving \NaD\ outflows, we appeal to ancillary radio data for our galaxies.   
In particular, we employ the \citet{Bes12} radio catalogue
of SDSS galaxies, where the presence of radio emission in each galaxy is  linked to either star-formation activity or a radio-AGN, using their stellar-population properties, the ratio of the
radio-to-optical-emission luminosity and standard BPT diagnostic
diagrams.
Fig.~\ref{fig:RAD_SY} again shows the $\Delta V$ vs. $\Delta \sigma$
diagram for our \NaD-detected Seyfert 2 galaxies, where the
coloured points now indicate whether the radio emission is dominated by star-formation or AGN activity. 
The \citeauthor{Bes12} catalog includes 97 (19\%) of our Seyfert 2 sample with interstellar \NaD\ absorption and, out of these objects, the radio emission is ascribed to a radio AGN only in 17 objects (18\%).
Most importantly Fig.~\ref{fig:RAD_SY} shows that, among the ten
radio-detected Seyfert 2 that also exhibit an \NaD-outflow, only one is classified as a radio AGN by
\citeauthor{Bes12}.
Radio-AGN Seyfert 2 galaxies, in fact, generally show nearly zero or
even slightly positive $\Delta V$ values, consistent with the results
of \citet{Sar16}, and also with what we find when cross-correlating the
\citeauthor{Bes12} catalogue with our control sample.

In summary, only a few dozen Seyfert 2 galaxies show kpc-scale
cold-gas outflows among the nearly complete sample of almost 10,000
SDSS galaxies that we have analysed. Furthermore, although nearly a
third of these outflowing systems feature some of the most luminous of
our Seyfert 2 nuclei, both the IR luminosity and the nature of the
radio emission in our Seyfert 2 galaxies indicates that star-formation
may be powering an even higher fraction of the \NaD-outflows observed in
the Seyfert 2 population. 

\section{Conclusions}

Using SDSS DR7 data and the value-added catalogues of
\citet{Oh11,Oh15}, we have selected a nearly complete sample of $\sim$10,000 nearby ($z<0.2$) Seyfert 2 galaxies, and investigated
the kinematics of their cold-gas medium at kpc-scales, as
traced by interstellar \NaD\ absorption lines. 
In order to set a detection threshold for possible cold-gas outflows and draw conclusions on the incidence of AGN-driven outflows in the nearby Universe, we have also measured the properties of the interstellar \NaD\ absorption in a carefully-selected sample of $\sim$44,000 control galaxies that match our Seyferts in redshift, luminosity, size, light concentration, apparent flattening, colour and morphological
classification. Our main results are as follows:

\begin{itemize}

\item The incidence of detected \NaD\ absorption across our Seyfert 2
  and control samples are similar, being 5.7\% and 4.9\% in the two populations respectively. This is particularly important, as it allows for a direct comparison between the cold-gas kinematics found across these two galaxy populations.

\item Out of 9859 Seyfert 2 galaxies, only 53 show evidence of
  kpc-scale cold-gas outflows. Even accounting for inclination biases
  against the detection of \NaD-outflowing systems, this result
  strongly suggests that optical AGN activity cannot be important in
  driving galactic winds capable of quenching star-formation in the
  nearby Universe.

\item The overall cold-gas kinematic behavior traced by the
  \NaD\ interstellar absorption in Seyfert 2 and control galaxies (as shown in the $V_{\rm Na\, D}-V_{\star}$ versus $\sigma_{\rm Na\,
    D}-\sigma_{\star}$ diagrams) is rather 
  similar, and the presence of an optical AGN does not
  boost the fraction of \NaD\ outflows compared to the control sample, where such outflows are likely driven by
  star-formation. In fact, the incidence of \NaD\ outflows
  among Seyfert 2s is actually lower, being 9\% of \NaD-detected objects, compared to 16\% for the control sample. 

\item Consistent with previous studies, many \NaD-outflowing Seyfert 2
  galaxies are some of the brightest AGN ($\log{L_{\rm
      [\textsc{O\,iii}]}} > 42.3\rm erg\,s^{-1} $). However, this only accounts for
  32\% of the outflows detected in Seyfert 2s.

\item On the other hand, ancillary radio and IR data of
  our Seyferts (available for 19\% and 55\%
  of the objects with interstellar \NaD\ absorption) suggests that
  star-formation is likely to be the bigger contributor to the observed
  \NaD-outflows. Among the \NaD-outflowing systems with radio
  or IR measurements, around 90\% show radio emission consistent with
  being powered by star formation rather than AGN activity, and 55\%
  show IR luminosity consistent with a LIRG or ULIRG
  classification. That a similar behaviour is observed across our
  control sample further indicates that star formation is the principal driver of many of the outflows observed in our Seyfert 2 galaxies.
\end{itemize}

The negligible fraction of Seyferts found to host outflows in this work reinforces the conclusions of recent studies \citep[e.g.][]{Kav15,Sar16}, which have suggested that the long time delay between the onset of star formation and the triggering of the AGN hinders the ability of the AGN to strongly regulate star formation activity. In essence, the gas reservoir is significantly depleted before the AGN is triggered, so that the AGN acts largely to mop up residual gas in the system, rather than influencing the bulk of the star formation episode. It is worth noting again that, since these papers have studied galaxies outside clusters, this scenario operates in low-density environments (which, nevertheless, host the vast majority of galaxies) at low redshift. As discussed in the introduction, the coupling between AGN and their hosts is stronger in clusters, and since performing similar \NaD\ analyses at high redshift is difficult, drawing conclusions about low-density environments in the early Universe is currently unfeasible.

While our results, combined with the recent literature, suggests that optical AGN activity may not play a significant role in quenching star formation in the nearby Universe, future integral-field observations, possibly assisted by adaptive optics, may shed further light on the relative role of AGN and star formation in powering the kpc-scale cold-gas outflows found in Seyfert 2 galaxies.
It is worth noting, in this context, that the [\textsc{O\,iii}] luminosity distribution of our Seyfert 2 sample extends to the more typical values observed in Seyfert 1 systems. Hence, even though the measurement of \NaD\ outflows
in these objects is generally hampered, either by the presence of a
non-thermal continuum, or by a real absence of neutral Sodium along the
line of sight, it appears unlikely that Seyfert 1s could power many more kpc-scale cold-gas outflows, just on the basis that their engines could possibly be more powerful than the AGN probed in this study. 
On the other hand, it also remains to be established how frequently the ionised-gas outflows often found in nearby Type 1 AGN \citep[e.g.,][]{Per17} really extend to kpc scales, again, for instance, through the use of integral-field spectroscopy \citep[e.g.,][]{Cre15}.

Integral-field observations may also clarify the impact of shocks on the nebular emission that is observed
in most of the \NaD-outflowing galaxies in our control sample, which
very often shows composite AGN/star-formating activity and LINER-like
emission. Indeed, if shocks do not always contribute to the observed
nebular spectrum \citep[as is the case in the study of ][]{Ho14}, this
may leave some room for AGN, possibly during a very obscured phase
\citep[][]{Per17,Harr17}, to contribute to the outflows observed in our control sample and also in the more general galaxy population.

A further avenue of exploration is the role of mergers in driving inflows and outflows of cold gas. Deep surveys like the Stripe 82 and, in particular, new datasets that are both wide and deep (e.g. DECaLS and those further downstream like LSST \citep{Ivez08, Abe09, Schl15, Blum16}) are capable of revealing faint tidal features due to recent interactions \citep[e.g.][]{Kav14b,Kav14a}. Such data can be combined with the \NaD\ analysis presented here to elucidate the role of merging in triggering galactic-scale outflows in the nearby Universe.

Nevertheless, to conclude, the results from this study suggest that galactic-scale outflows at low redshift are no more frequent in Seyferts than they are in similar non-active galaxies, that optical AGN are not significant contributors to the quenching of star formation in the nearby Universe and that star formation may be the principal driver of outflows even in systems that host an AGN. 

\section{ACKNOWLEDGEMENTS}
Both BN and MS acknowledge the financial support and hospitality provided by the European Southern Observatory (ESO) through the award of an ESO studentship and visiting fellowship, respectively. SK acknowledges a Senior Research Fellowship from Worcester College Oxford. This work would not have been possible without the data taken from the Sloan Digital Sky Survey (SDSS), the SDSS added-value OSSY  emission-line catalogue and the contribution of large numbers of citizen scientists towards classifying the morphology of SDSS galaxies.

%\clearpage

\bsp
\label{lastpage}
\end{document}